\documentclass[twocolumn,aps,superscriptaddress]{revtex4}
\usepackage{amssymb}
\usepackage{amsmath}
\usepackage{graphicx}
\usepackage[normalem]{ulem}
\usepackage[dvips]{color}
\usepackage{appendix}

\begin{document}

\title{Constraining isovector nuclear interactions with giant dipole resonance and neutron skin in $^{208}$Pb from a Bayesian approach}

\author{Jun Xu\footnote{xujun@zjlab.org.cn}}
\affiliation{The Interdisciplinary Research Center, Shanghai Advanced Research Institute, Chinese Academy of Sciences, Shanghai 201210, China}
\affiliation{Shanghai Institute of Applied Physics, Chinese Academy of Sciences, Shanghai 201800, China}

\begin{abstract}
The remaining uncertainties of isovector nuclear interactions call for reliable experimental measurements of isovector probes in finite nuclei. Based on the Bayesian analysis, although the neutron-skin thickness data or the isovector giant dipole resonance data in $^{208}$Pb can constrain only one isovector interaction parameter, correlations between other parameters are built. Using combined data of both the neutron-skin thickness and the isovector giant dipole resonance helps to constrain significantly all isovector interaction parameters, thus serves as a useful way in the future analysis.
\end{abstract}

\maketitle
Extracting properties of nuclear interactions from observables in finite nuclei is an effective way in understanding the strong interaction with less uncertainties. Thanks to pioneering studies by nuclear physicists, isoscalar nuclear interactions are so far better constrained, while larger uncertainties still exist mainly in isovector nuclear interactions, and this hampers us from understanding more accurately properties of radioactive nuclei, dynamics in nuclear reactions induced by neutron-rich nuclei, and many interesting astrophysical phenomena such as the gravitational wave emerging from neutron-star mergers. On one hand, isovector nuclear interactions manifest itself in the isospin-dependent part of the nuclear matter equation of state, i.e., the nuclear symmetry energy $E_{\rm{sym}}$~\cite{Bar05,Ste05,Lat07,Li08}, which is generally characterized by its value $E_{\rm{sym}}^0$ at the saturation density $\rho_0=0.16$ fm$^{-3}$ as well as its slope parameter $L=3\rho_0(dE_{\rm{sym}}/d\rho)_{\rho_0}$ around the saturation density. In recent years, constraints of $E_{\rm{sym}}(\rho_0)=31.7 \pm 3.2$~MeV and $L = 58.7 \pm 28.1$~MeV were obtained from surveying 53 analyses carried out by 2016 using various terrestrial nuclear laboratory data and astrophysical observations~\cite{BAL13,Oer17}, while more accurate constraints are still called for. On the other hand, isovector nuclear interactions also determine properties of single-nucleon potentials in neutron-rich medium. As a result of the momentum-dependent single-nucleon potential, nucleons move with an effective mass $m^\star$ instead of a free mass $m$ in nuclear matter, similar to the case of electron dynamics near the gap between energy bands in semiconductors. Different neutron and proton effective masses affect not only isospin-dependent nucleon dynamics in heavy-ion collisions but also properties of neutron-rich nuclei~\cite{Li18}. Since both the nuclear symmetry energy $E_{\rm{sym}}$ and the neutron-proton effective mass splitting $m^\star_{n-p}=m^\star_n-m^\star_p$ originate from isovector nuclear interactions, it is not surprising that they are related to each other through the Hugenholtz-Van Hove theorem~\cite{BAL13,XuC10}.

The present study serves as a trial to constrain isovector nuclear interactions manifested by the nuclear symmetry energy and the neutron-proton effective mass splitting from properties of $^{208}$Pb, a heavy spherical nucleus with less ambiguities of clustering, deformation, etc. The main focuses are on the isovector giant dipole resonance (IVGDR) and the neutron-skin thickness in $^{208}$Pb. The isovector giant dipole resonance can be considered as a collective excitation mode with neutrons and protons moving relatively to each other in a nucleus like a harmonic oscillator, where the symmetry energy acts as an restoring force~\cite{Tri08,Rei10,Pie12,Vre12,Roc13b,Col14,Roc15,zhangzhen15,zhenghua16,Xu20a} while the nucleon effective mass, which is analogue of the oscillator mass, may also affect the dynamics~\cite{Zha16,Kon17}, so the deexcitation spectrum measured experimentally can be a probe of both of them. The neutron-skin thickness, defined as the difference in neutron and proton radii in a nucleus, is mostly caused by the stronger pressure for neutrons than that for protons, and is one of the most robust and sensitive probes of the symmetry energy slope parameter~\cite{Bro00,Typ01,Chuck01,Fur02,Tod05,Cen09,Zha13,India1,India2,Thiel,Burg,Vin14,X18}. Although each observable is not expected to pin down both the symmetry energy and the neutron-proton effective mass splitting, we will show that using both observables in the same $^{208}$Pb nucleus helps to constrain significantly isovector nuclear interactions.

In the theoretical calculation of the isovector giant dipole resonance and neutron-skin thickness, we use the Skyrme-Hartree-Fock (SHF) model originating from the following effective interaction between two nucleons at $\vec{r}_1$ and $\vec{r}_2$
\begin{eqnarray}\label{SHFv}
v(\vec{r}_1,\vec{r}_2) &=& t_0(1+x_0P_\sigma)\delta(\vec{r}) \notag \\
&+& \frac{1}{2} t_1(1+x_1P_\sigma)[{\vec{k}'^2}\delta(\vec{r})+\delta(\vec{r})\vec{k}^2] \notag\\
&+&t_2(1+x_2P_\sigma)\vec{k}' \cdot \delta(\vec{r})\vec{k} \notag\\
&+&\frac{1}{6}t_3(1+x_3P_\sigma)\rho^\alpha(\vec{R})\delta(\vec{r}) \notag\\
&+& i W_0(\vec{\sigma}_1+\vec{\sigma_2})[\vec{k}' \times \delta(\vec{r})\vec{k}].
\end{eqnarray}
In the above, $\vec{r}=\vec{r}_1-\vec{r}_2$ and $\vec{R}=(\vec{r}_1+\vec{r}_2)/2$ are respectively the relative and the central coordinate, $\vec{k}=(\nabla_1-\nabla_2)/2i$ is the relative momentum operator and $\vec{k}'$ is its complex conjugate acting on the left, and $P_\sigma=(1+\vec{\sigma}_1 \cdot \vec{\sigma}_2)/2$ is the spin exchange operator, with $\vec{\sigma}_{1(2)}$ being the Pauli matrics. The spin-orbit coupling constant is fixed at $W_0=133.3$ MeVfm$^5$. Instead of the usual fitting process, the other 9 SHF parameters $t_0$, $t_1$, $t_2$, $t_3$, $x_0$, $x_1$, $x_2$, $x_3$, and $\alpha$ can be inversely solved from 9 macroscopic quantities characterizing saturation properties of nuclear matter~\cite{MSL0}, among which are the isoscalar and isovector nucleon effective mass $m_s^\star$ and $m_v^\star$ at the Fermi momentum in normal nuclear matter, and the symmetry energy $E_{sym}^0$ and its slope parameter $L$ at the saturation density. The isoscalar nucleon effective mass $m_s^\star$ is the nucleon effective mass in isospin symmetric nuclear matter, while the isovector nucleon effective mass $m_v^\star$ is the proton (neutron) effective mass in pure neutron (proton) matter. Up to the linear order of the isospin asymmetry $\delta$, the neutron-proton effective mass splitting is related to the $m_s^\star$ and $m_v^\star$ through the relation
\begin{eqnarray}\label{mnp}
m_n^*-m_p^* \approx \frac{2m_s^*}{m_v^*}(m_s^*-m_v^*)\delta.
\end{eqnarray}
Determining the SHF parameters from expressions of macroscopic quantities~\cite{MSL0} provides the possibility of changing the value of only one of them while keeping the values of others the same, so in this way one can investigate the effect on observables from only one particular macroscopic quantity. In the present study, we fix $m_s^\star=0.83m$ in order to reproduce the excitation energy of the isoscalar giant quadruple resonance approximately independent of other macroscopic quantities, and change only one individual quantity $E_{sym}^0$, $L$, or $m_v^\star$ at each time, while keeping the values of other macroscopic quantities as the empirical values as in Table I of Ref.~\cite{MSL0}.

Based on the Hartree-Fock method, the effective interaction [Eq.~(1)] leads to the standard SHF energy-density functional~\cite{Cha97}, with time-odd terms neglected in the calculation of spin-saturated spherical nuclei. With the single-nucleon Hamiltonian obtained via the variational principle from the energy-density functional~\cite{Vau72}, the nucleon wave functions can be calculated from the Schr\"odinger equation, and this gives the nucleon density distributions. The Reinhard's SHF code~\cite{SHFcode} is used to calculate the nucleon density distribution and the neutron-skin thickness $\Delta r_{np}$ in $^{208}$Pb from the above standard procedure.

With the nucleon wave functions obtained from the Hartree-Fock method, the random-phase approximation (RPA) method can be used to calculate the strength function
\begin{equation}
S(E) = \sum_\nu |\langle \nu|| \hat{F}_{J}  || \tilde{0} \rangle |^2 \delta(E-E_\nu)
\end{equation}
of a particular nucleus resonance, where the square of the reduced matrix element $|\langle \nu|| \hat{F}_{J}  || \tilde{0} \rangle |$ represents the transition probability from the ground state $| \tilde{0} \rangle $ to the excited state $| \nu \rangle$, with $\hat{F}_{J}$ being the operator of the nucleus excitation. For IVGDR, the operator can be expressed as
\begin{equation}
\hat{F}_{\rm 1M} = \frac{N}{A} \sum_{i=1}^Z r_i Y_{\rm 1M}(\hat{r}_i) - \frac{Z}{A} \sum_{i=1}^N r_i Y_{\rm 1M}(\hat{r}_i), \label{QIVGDR}
\end{equation}
where $N$, $Z$, and $A$ are respectively the neutron, proton, and nucleon numbers in a nucleus, $r_i$ is the coordinate of the $i$th nucleon with respect to the center-of-mass of the nucleus, and $Y_{\rm 1M}(\hat{r}_i)$ is the spherical Bessel functions, with the magnetic quantum number $M$ degenerate in spherical nuclei. The two observables charaterizing the strength function of IVGDR, i.e., the centroid energy $E_{-1}$ and the electric polarizability $\alpha_D$, can be obtained from the moments of the strength function
\begin{equation}
m_k = \int_0^\infty dE E^k S(E)
\end{equation}
through the relations
\begin{equation}
E_{-1} = \sqrt{m_1/m_{-1}}, ~\alpha_D = \frac{8\pi e^2}{9} m_{-1}.
\end{equation}
The open source code developed in Ref.~\cite{Col13} is used to calculate the $E_{-1}$ and $\alpha_D$ of the IVGDR in $^{208}$Pb using the RPA method based on the SHF model.

We have employed the Bayesian analysis method to obtain the probability distribution function (PDF) of the symmetry energy $E_{\rm{sym}}$ and the isovector nucleon effective mass $m^\star_v$ from the neutron-skin thickness $\Delta r_{np}$ as well as the centroid energy $E_{-1}$ and the electric polarizability $\alpha_D$ in $^{208}$Pb. How the experimental data improves our knowledge of model parameters can be described by the Bayes' theorem formally written as
\begin{equation}\label{Bayes}
P(M|D) = \frac{P(D|M)P(M)}{\int P(D|M)P(M)dM}.
\end{equation}
In the above, $P(M|D)$ is the posterior PDF for the model $M$ given the experimental data $D$, which is seen to be normalized by the denominator of the right-hand side. $P(M)$ denotes the prior PDF of the model $M$ before being confronted with the experimental data, and we choose the prior PDFs of the model parameters $p_1=L$ uniformly distributed within $(0,120)$ MeV, $p_2=E_{sym}^0$ uniformly distributed within $(25,35)$ MeV, and $p_3=m^\star/m$ uniformly distributed within $(0.5,1)$. In this way, we investigate what information can $\Delta r_{np}$ and/or IVGDR alone can provide on the model parameters, while using prior PDFs from other studies may introduce additional information on the posterior PDFs of these parameters. Within these large prior distribution ranges of isovector interaction parameters, the binding energies and charge radii of $^{208}$Pb are deviated by only a few percent at most~\cite{MSL0}, showing that we are exploring the reasonable parameter space. $P(D|M)$ is the likelihood function describing how well the theoretical model $M$ predicts the experimental data $D$, and it is defined as
\begin{equation}
P[D(d_{1,2,3})|M(p_{1,2,3})] = \Pi_{i} \frac{1}{\sqrt{2\pi} \sigma_i}\exp\left[-\frac{(d_i^{exp}-d_i^{theo})^2}{2\sigma_i^2}\right], \label{llh}
\end{equation}
where $d_{1,2,3}$ represent the experimental data of $\Delta r_{np}$, $E_{-1}$, and $\alpha_D$ in $^{208}$Pb, respectively, and $d_i^{theo}$ and $d_i^{exp}$ are respectively the theoretical result and the mean value of the experimental result of the $i$th observable. In principle, $\sigma_i$ denoting the width of the likelihood function represents both the experimental and theoretical error, while it is chosen as the $1\sigma$ error bar of the experimental measurement in the present study. For the IVGDR in $^{208}$Pb, the centroid energy $E_{-1}=13.46$ MeV was accurately measured from photoneutron scatterings~\cite{IVGDRe}, and the electric polarizability $\alpha_D=19.6 \pm 0.6$ fm$^3$ was obtained from polarized proton inelastic scatterings~\cite{Tam11} and with the quasi-deuteron excitation contribution subtracted~\cite{Roc15}. An artificial error bar of 0.1 MeV is used for the well-determined $E_{-1}$ value of the IVGDR. For the neutron-skin thickness in $^{208}$Pb, the values of $\Delta r_{np}$ are different from different measurements. For example, $\Delta r_{np}=0.211^{+0.054}_{-0.063}$ fm and $0.16\pm0.07$ fm were obtained from proton~\cite{Zen10} and pion~\cite{Fri12} scatterings, respectively, $\Delta r_{np}=0.18\pm 0.04({\rm expt.}) \pm 0.05({\rm theor.})$ fm was obtained from the annihilation of antiprotons on the nuclear surface~\cite{Klo07,Bro07}, $\Delta r_{np}=0.15 \pm 0.03 ({\rm stat.}) ^{+0.01}_{-0.03} ({\rm sys.})$ fm was obtained from coherent pion photoproductions~\cite{Tar14}, and $\Delta r_{np}=0.33^{+0.16}_{-0.18}$ fm was measured from parity-violating electron scatterings~\cite{Abr12}. In the present study, we use the imagined experimental data of $\Delta r_{np}=0.15$ and 0.20 fm, each with imagined error bars of 0.02 and 0.06 fm, and meanwhile waiting for the new data by the lead ($^{208}$Pb) radius experiment at the Jeferson Laboratory and the Mainz radius experiment at the Mainz energy recovery superconducting accelerator~\cite{Thiel}. The calculation of the posterior PDFs is based on the Markov-Chain Monte Carlo approach using the Metropolis-Hastings algorithm, and the samples are analyzed after convergence without the initial burn-in steps. The PDF of a single model parameter $p_i$ is given by
\begin{equation}\label{1dpdf}
P(p_i|D) = \frac{\int P(D|M) P(M) \Pi_{j\ne i} dp_j}{\int P(D|M) P(M) \Pi_{j} dp_j},
\end{equation}
while the correlated PDF of two model parameters $p_i$ and $p_j$ is given by
\begin{equation}\label{2dpdf}
P[(p_i,p_j)|D] = \frac{\int P(D|M) P(M) \Pi_{k\ne i,j} dp_{k}}{\int P(D|M) P(M) \Pi_{k} dp_k}.
\end{equation}

\begin{figure}[ht]
	\centering\includegraphics[scale=0.12]{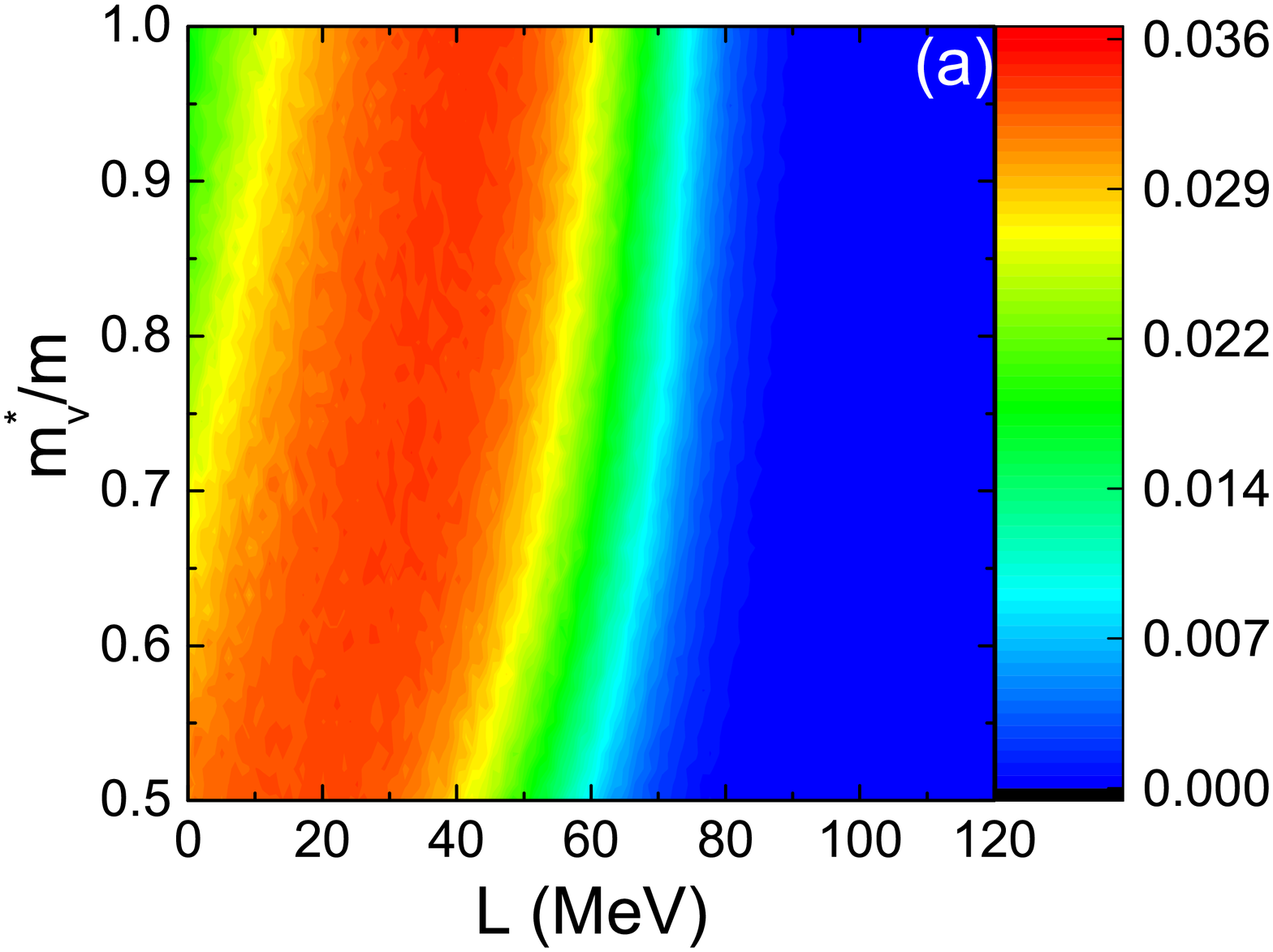}\includegraphics[scale=0.12]{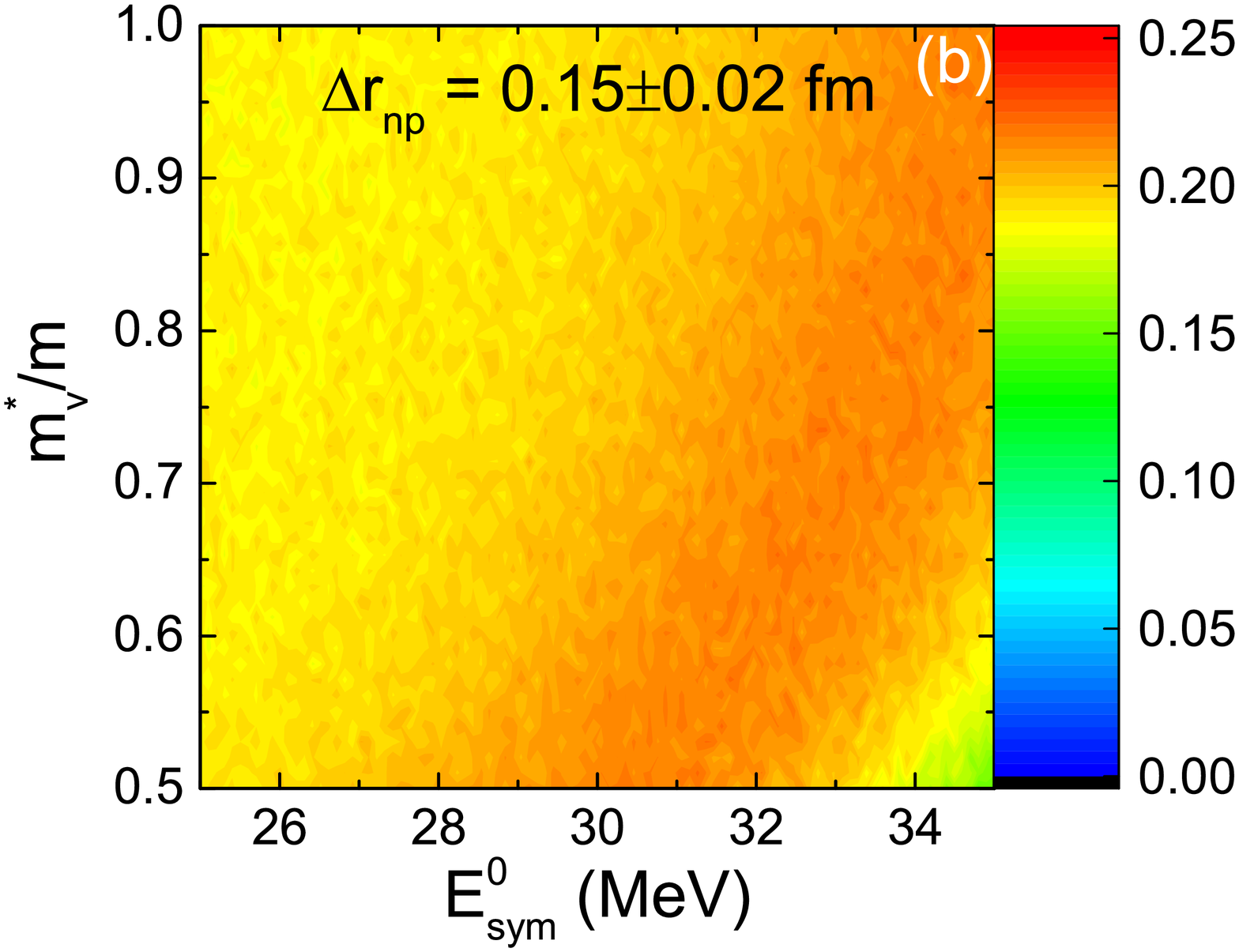}\includegraphics[scale=0.12]{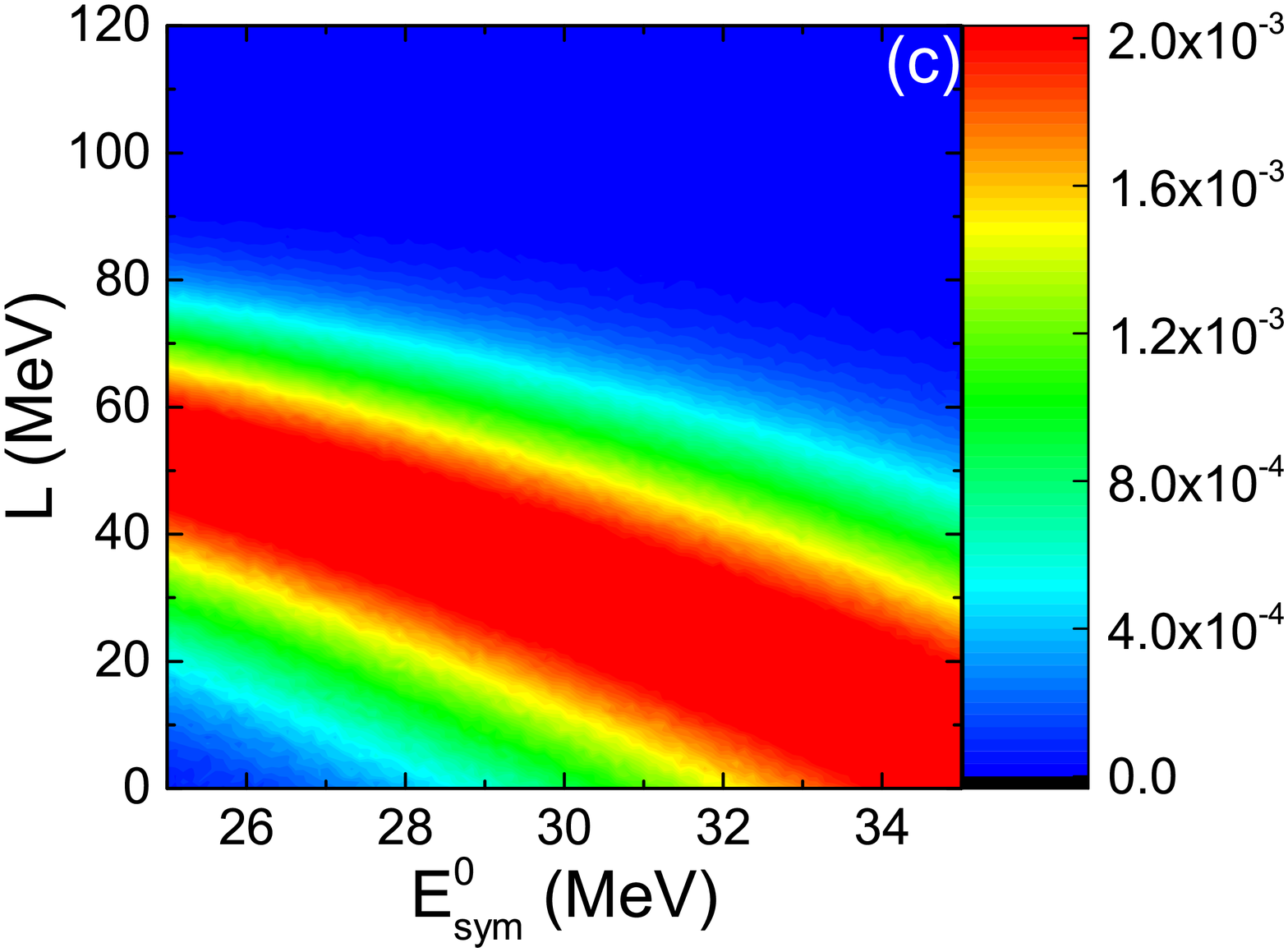}\\
	\centering\includegraphics[scale=0.12]{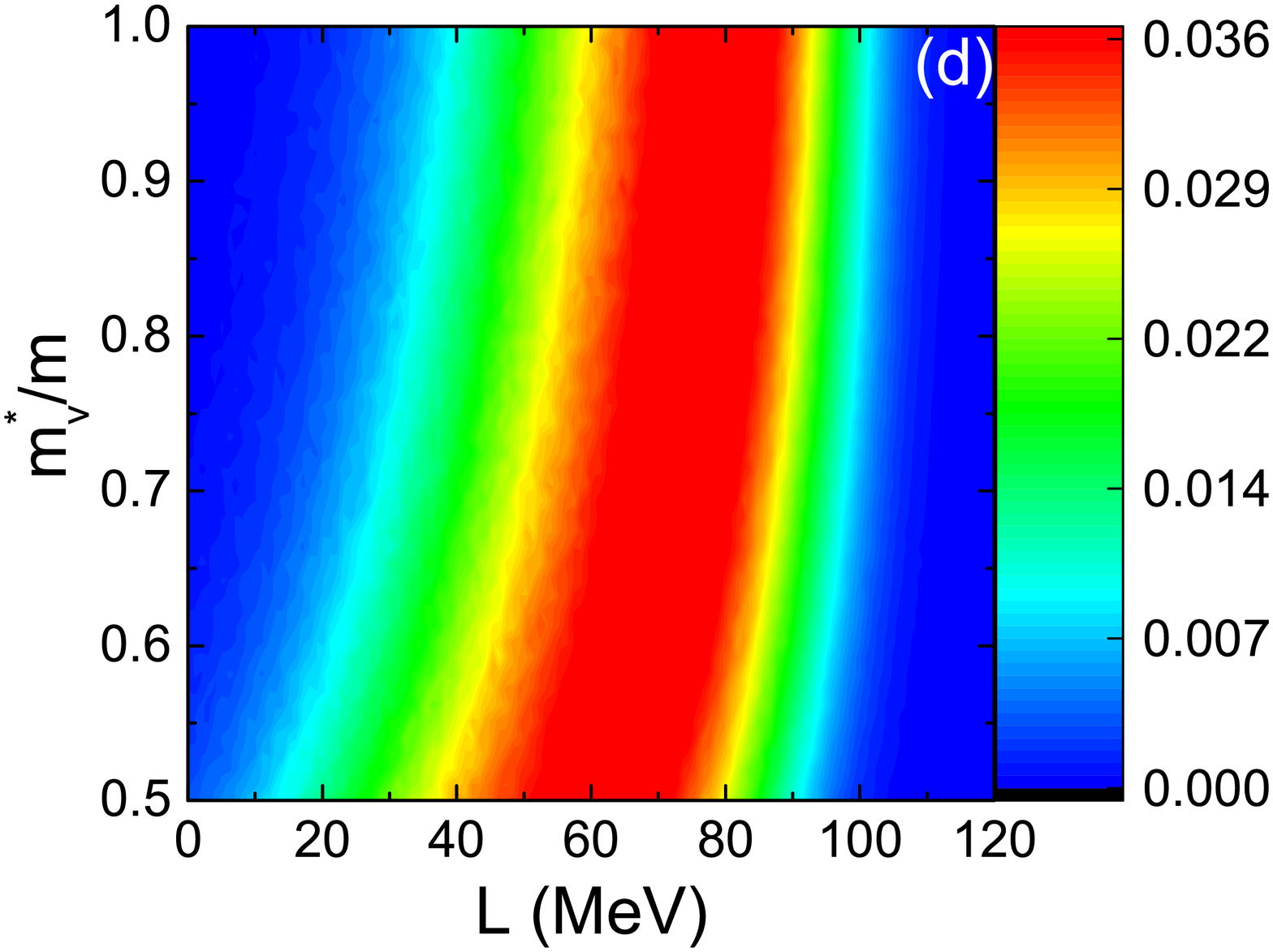}\includegraphics[scale=0.12]{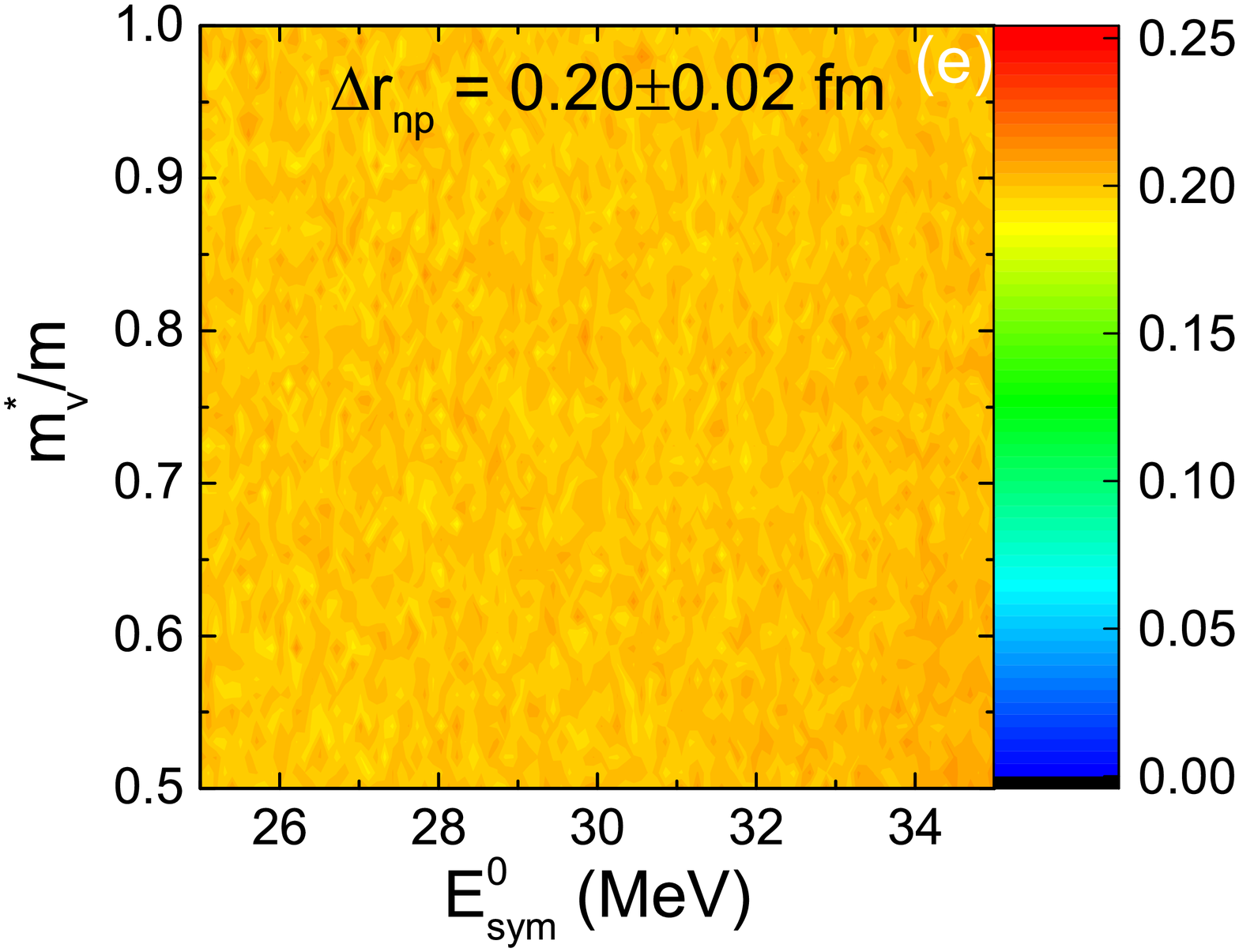}\includegraphics[scale=0.12]{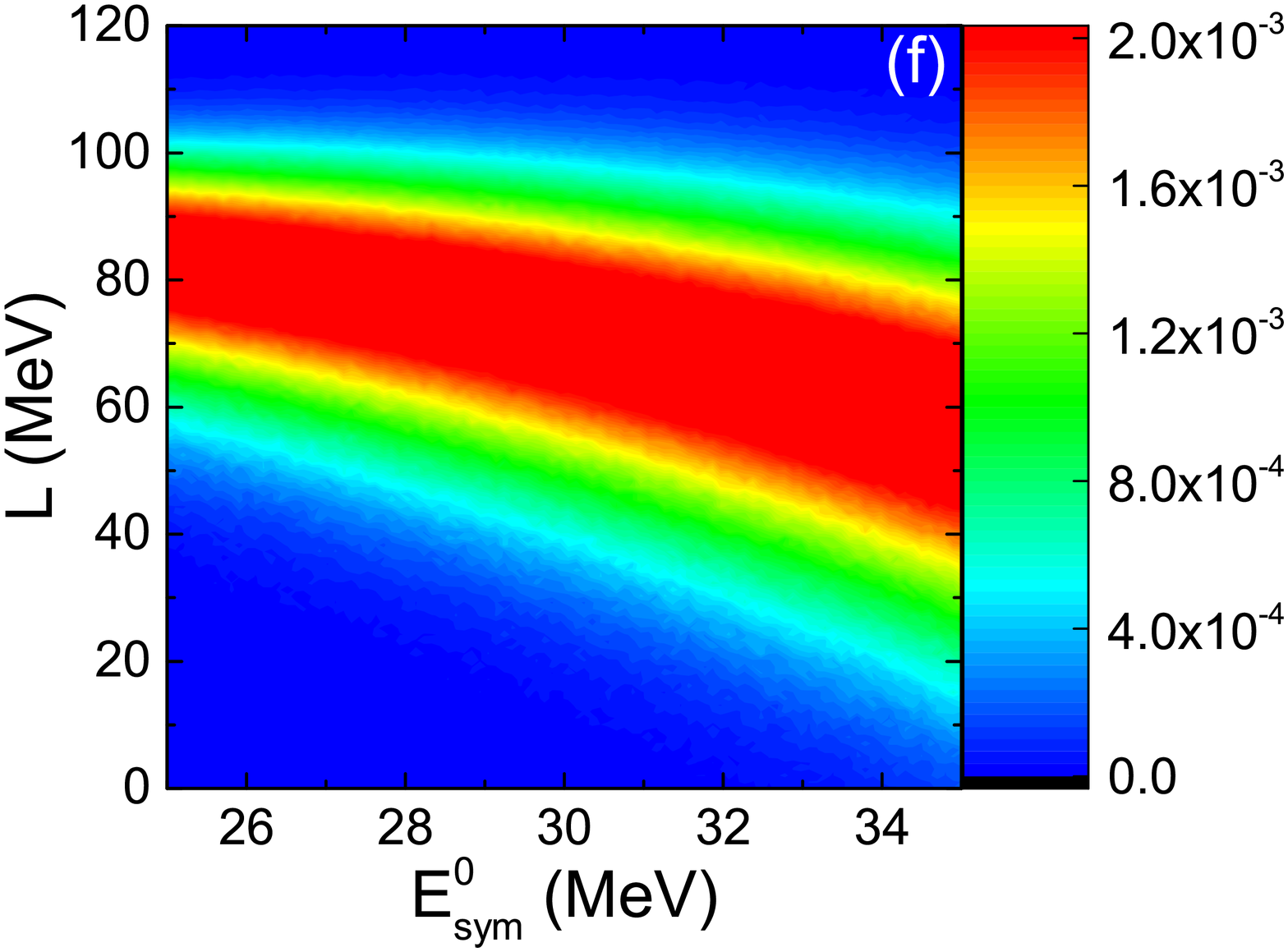}\\
	\centering\includegraphics[scale=0.12]{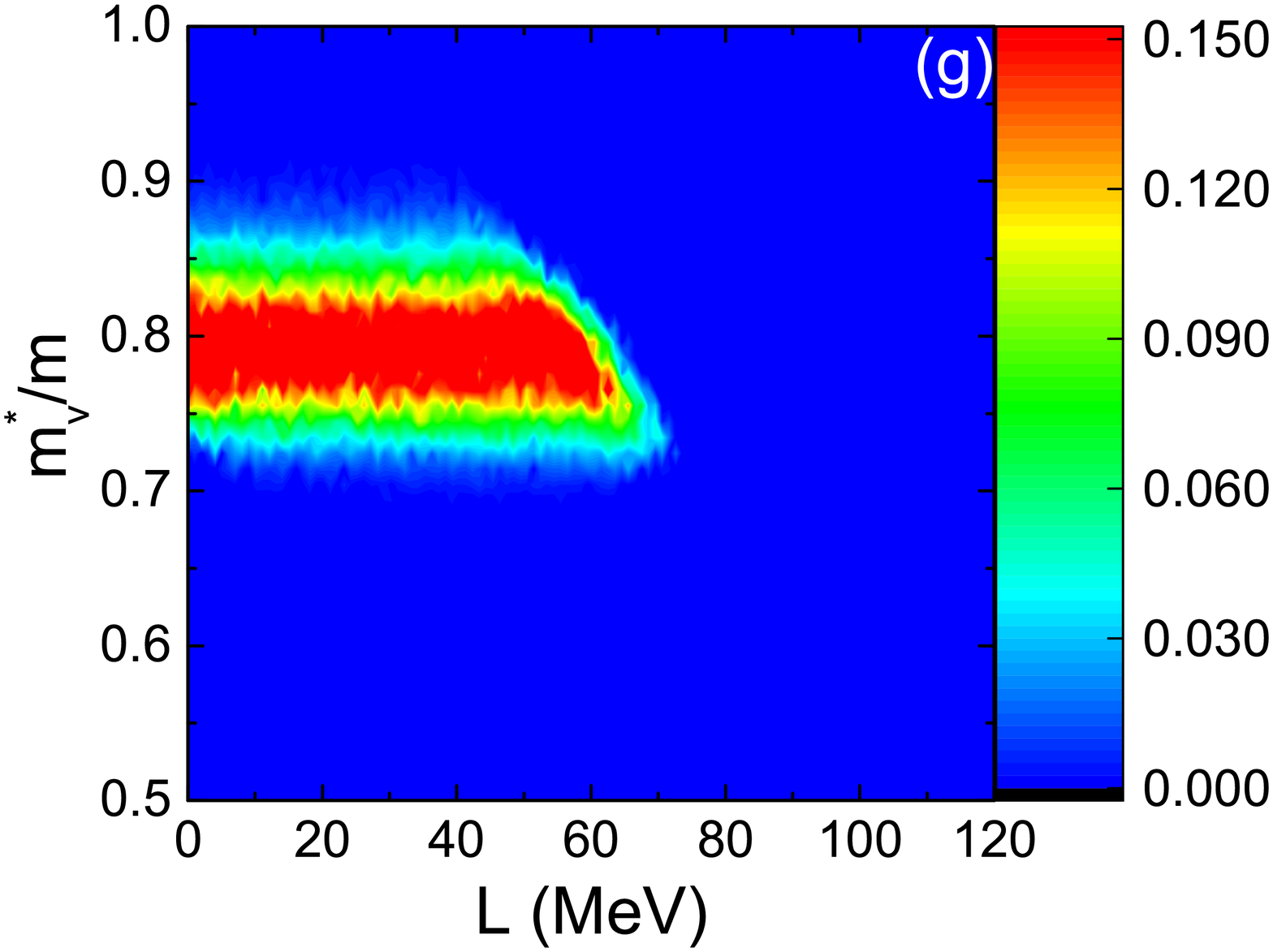}\includegraphics[scale=0.12]{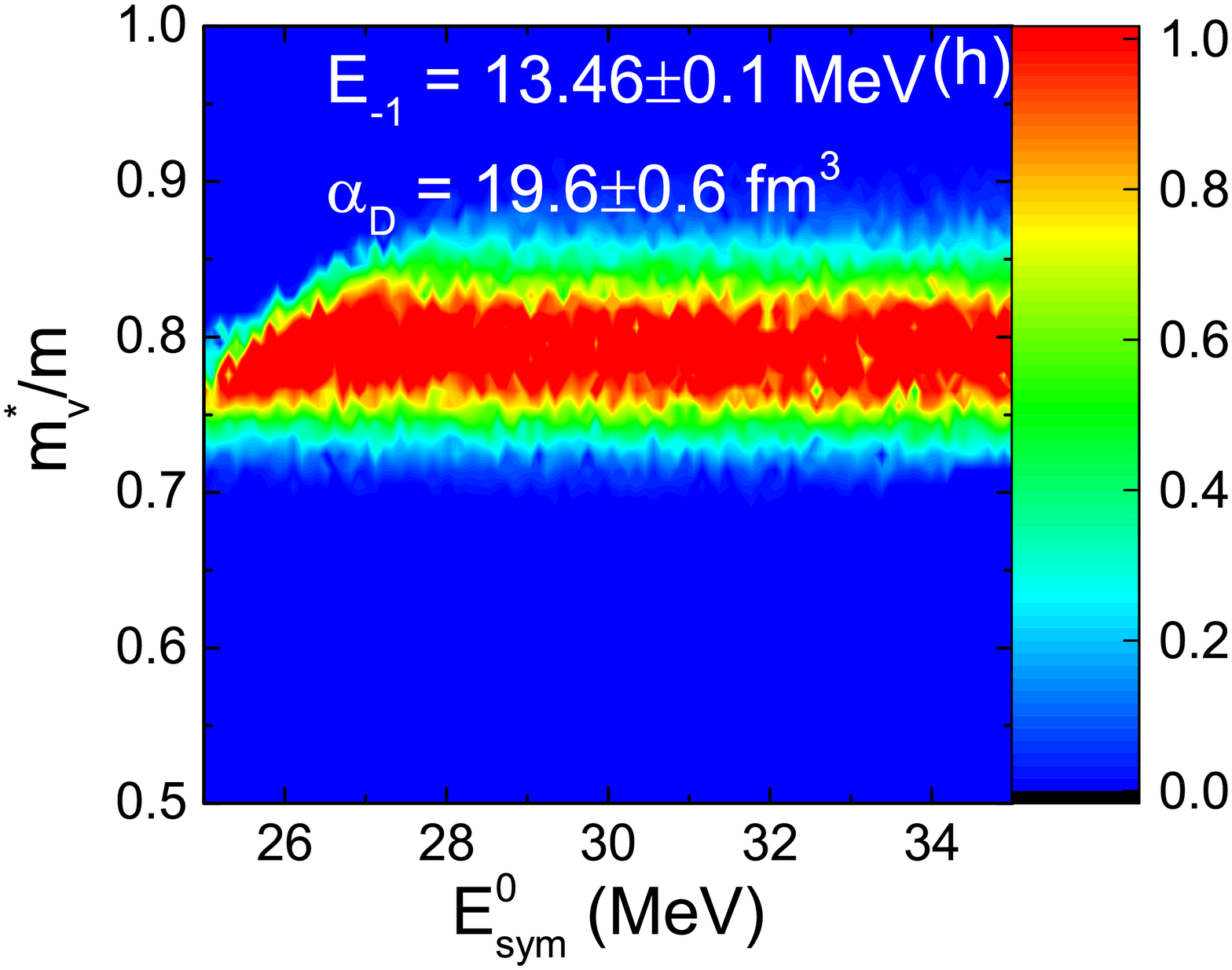}\includegraphics[scale=0.12]{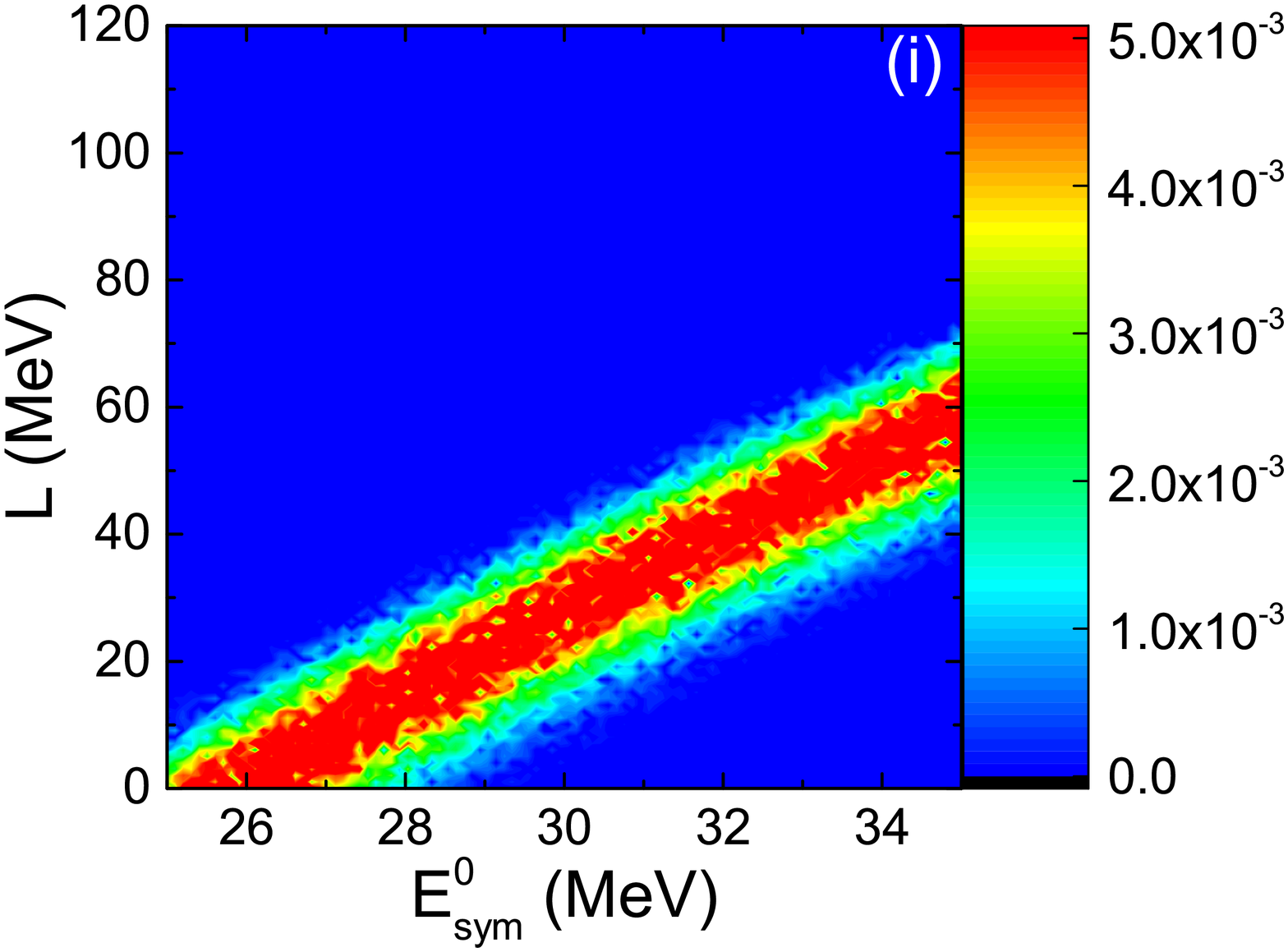}
	\caption{(Color online) Correlated posterior PDFs from the neutron-skin thickness data of $\Delta r_{np}=0.15\pm0.02$ fm (upper panels), $\Delta r_{np}=0.20\pm0.02$ fm (middle panels), or the IVGDR data (bottom panels) in $^{208}$Pb in the $L-m_v^\star/m$ plane (left), the $E_{sym}^0-m_v^\star/m$ plane (middle), and the $E_{sym}^0-L$ plane (right).} \label{fig1}
\end{figure}

\begin{figure}[ht]
	\centering\includegraphics[scale=0.35]{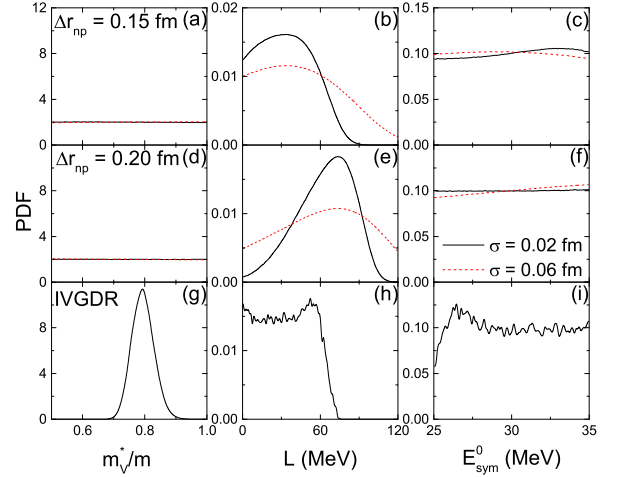}
	\caption{(Color online) Posterior PDFs from the neutron-skin thickness data of $\Delta r_{np}=0.15$ fm (upper panels) or $\Delta r_{np}=0.20$ fm (middle panels) with error bars 0.02 and 0.06 fm, or the IVGDR data (bottom panels) in $^{208}$Pb for $m_v^\star/m$ (left), $L$ (middle), and $E_{sym}^0$ (right).} \label{fig2}
\end{figure}

\begin{figure}[ht]
	\centering\includegraphics[scale=0.12]{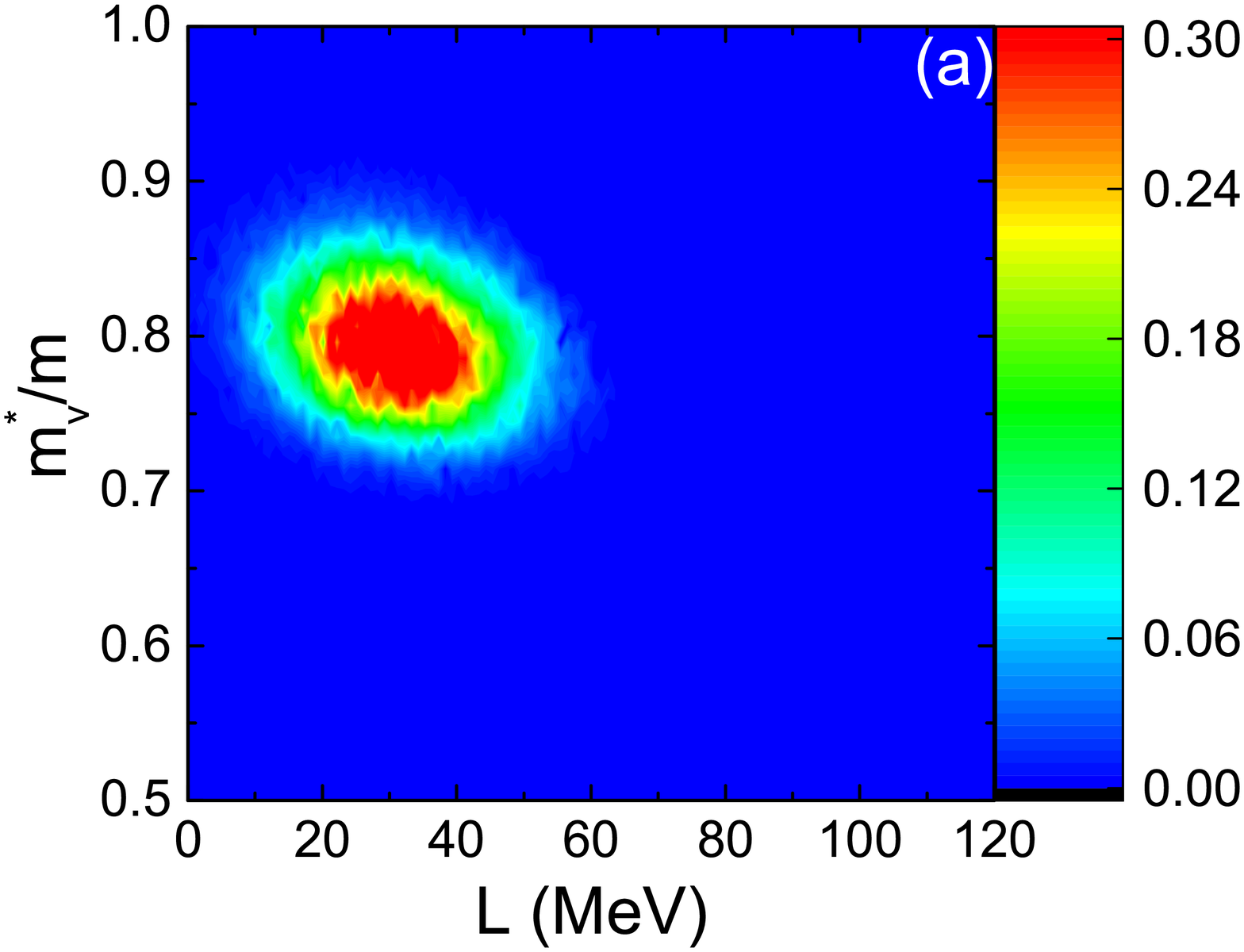}\includegraphics[scale=0.12]{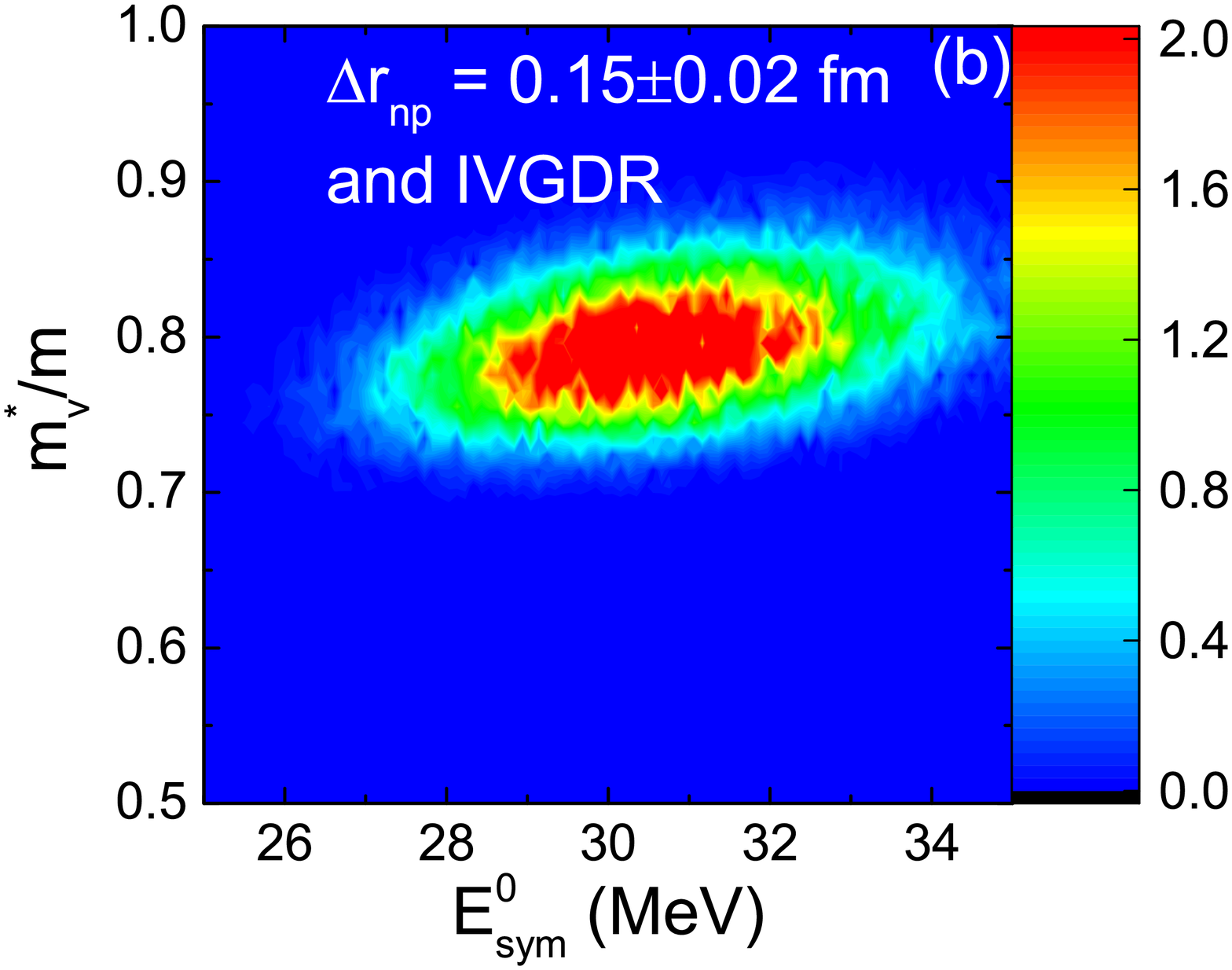}\includegraphics[scale=0.12]{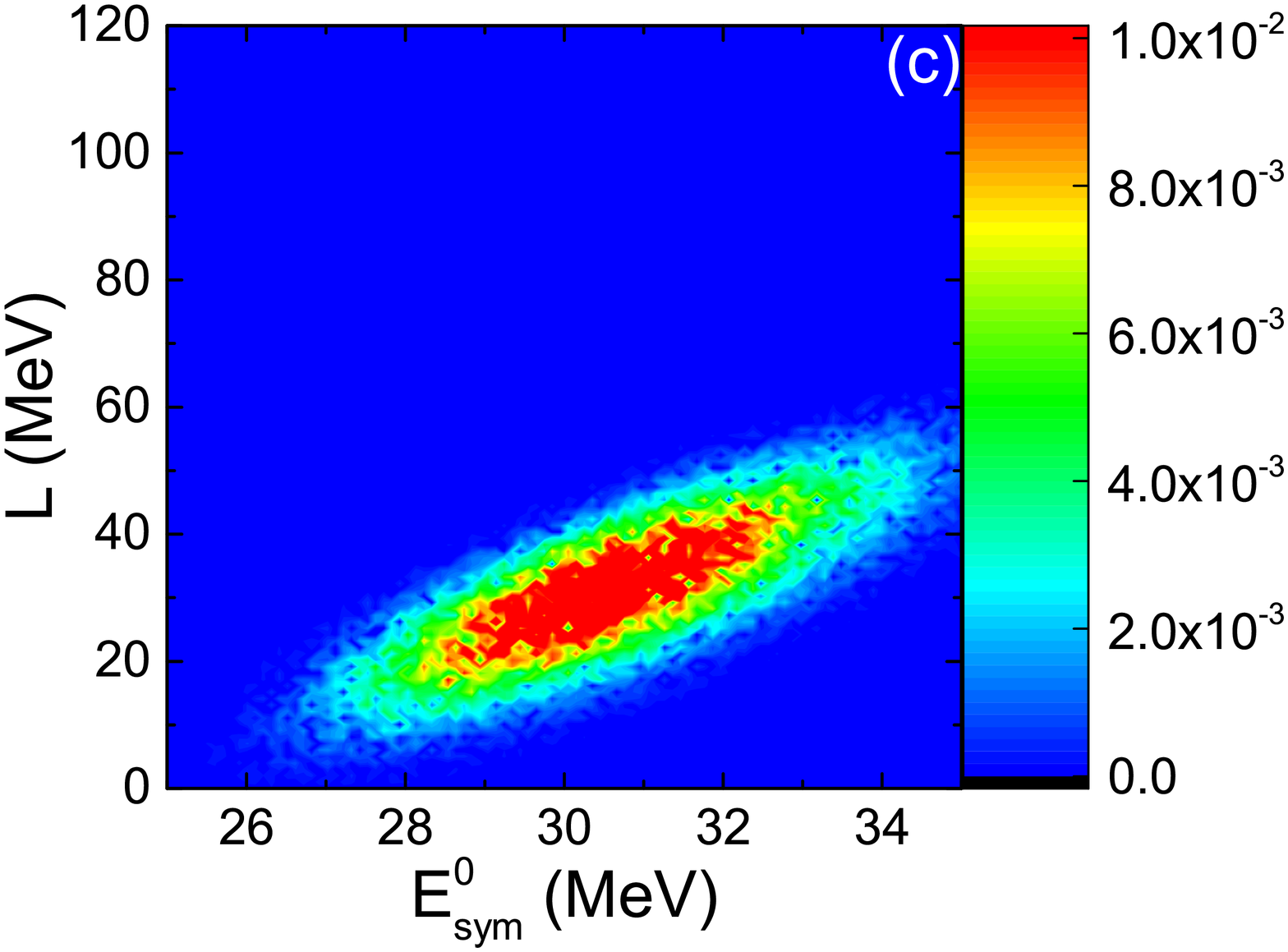}\\
	\centering\includegraphics[scale=0.12]{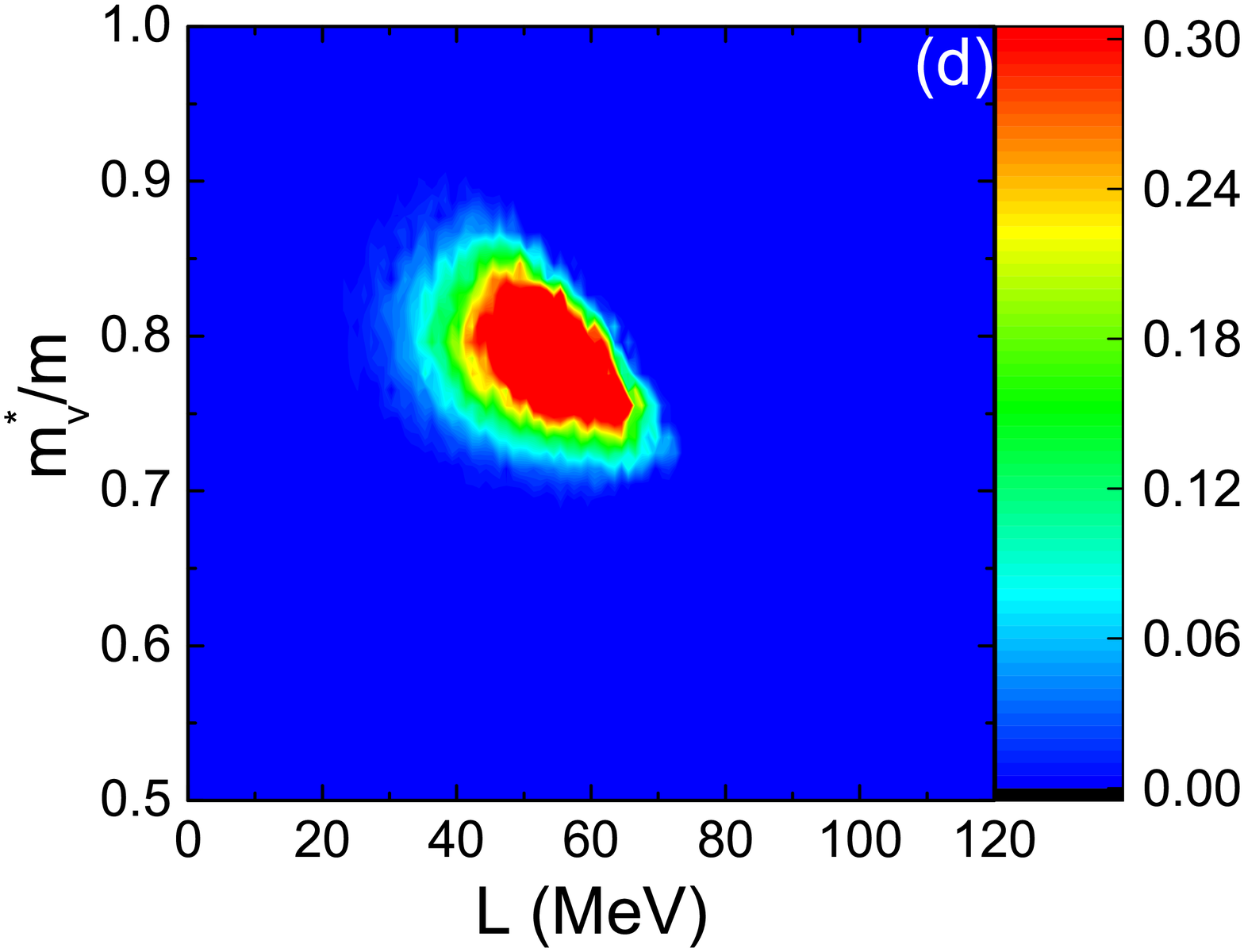}\includegraphics[scale=0.12]{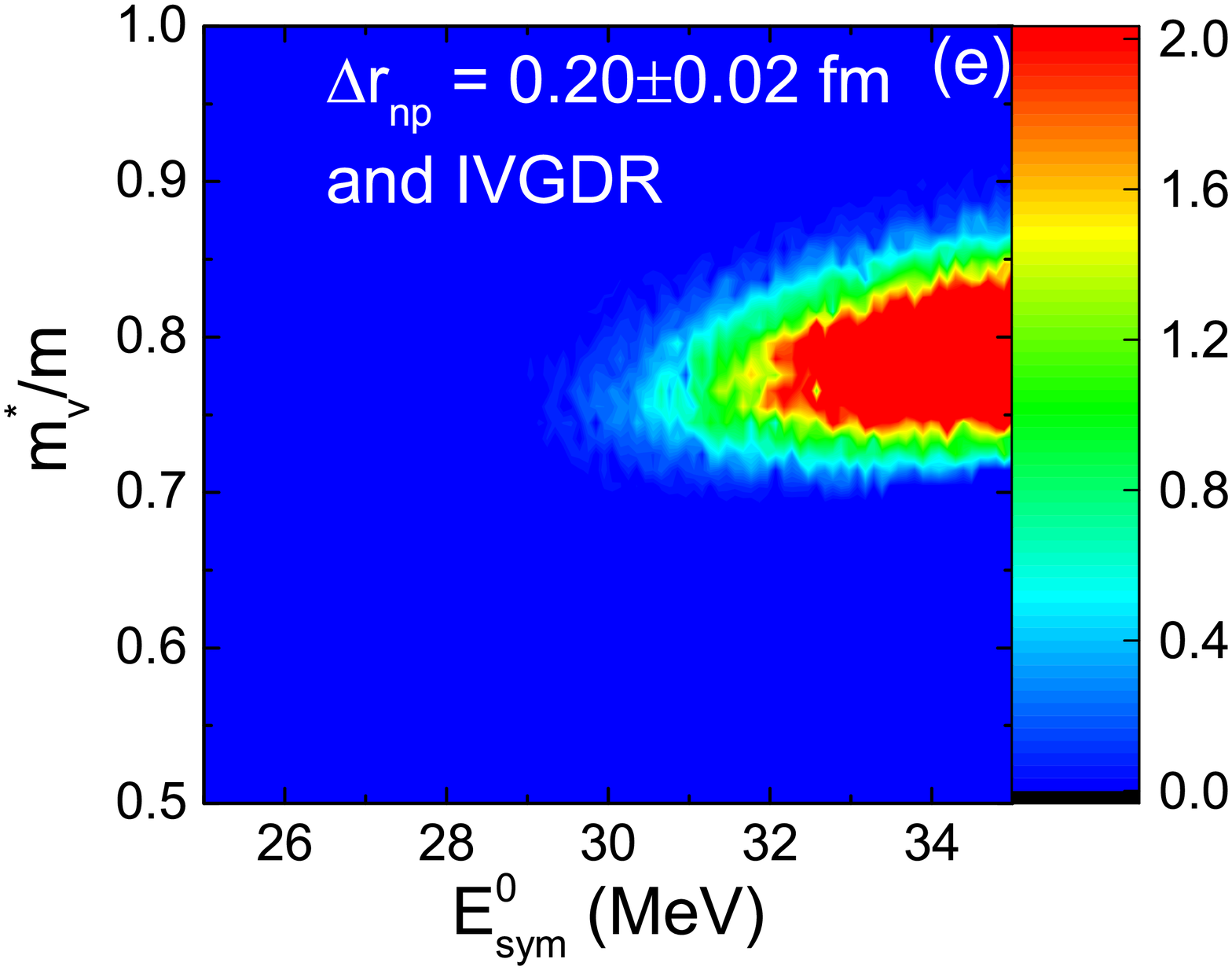}\includegraphics[scale=0.12]{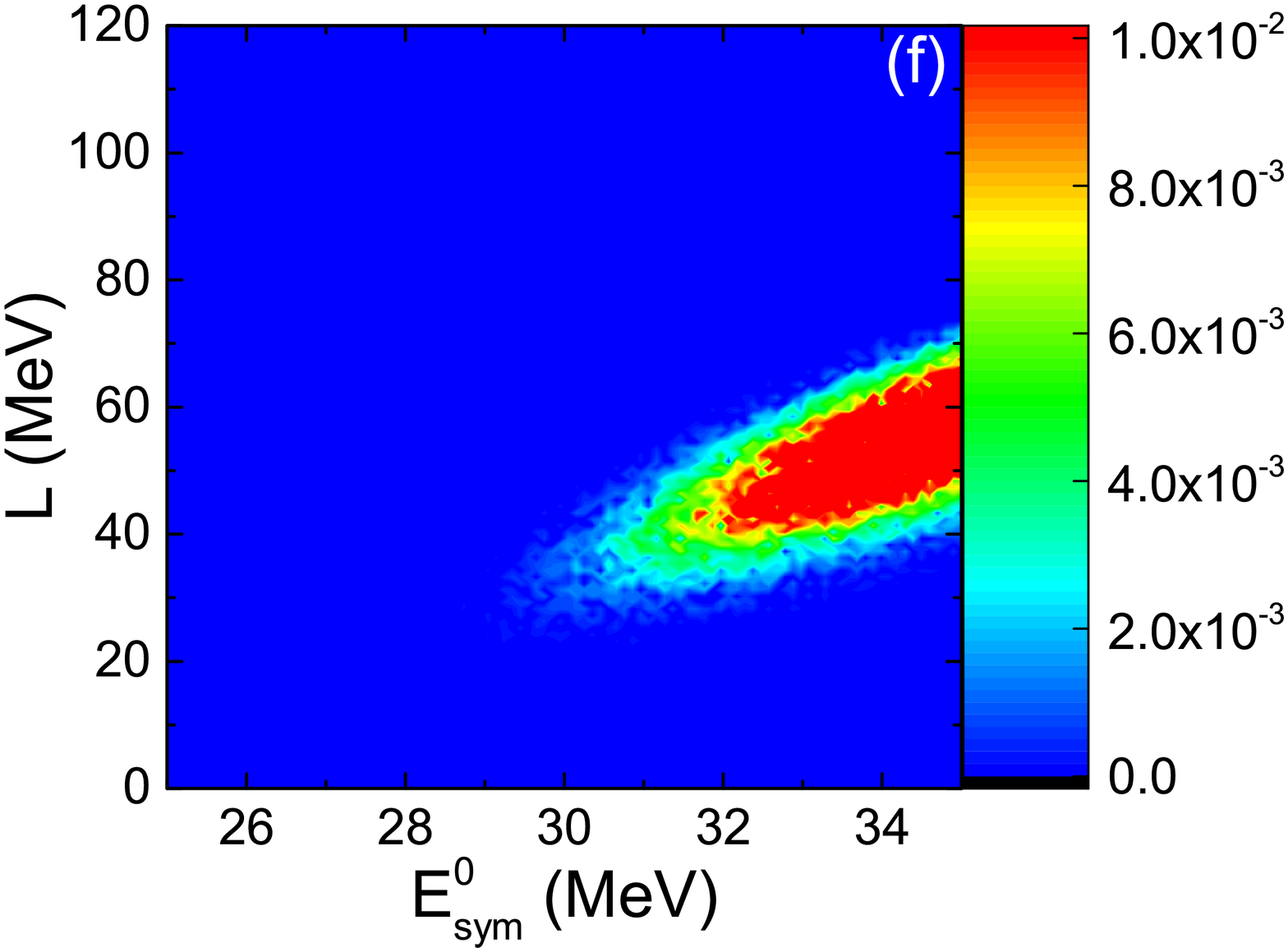}
	\caption{(Color online) Correlated posterior PDFs from combining the neutron-skin thickness data [$\Delta r_{np}=0.15\pm0.02$ fm (upper panels) or $\Delta r_{np}=0.20\pm0.02$ fm (lower panels)] and the IVGDR data in $^{208}$Pb in the $L-m_v^\star/m$ plane (left), the $E_{sym}^0-m_v^\star/m$ plane (middle), and the $E_{sym}^0-L$ plane (right).} \label{fig3}
\end{figure}

\begin{figure}[ht]
	\centering\includegraphics[scale=0.35]{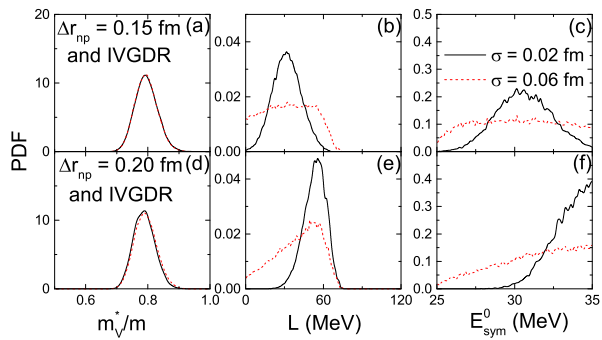}
	\caption{(Color online) Posterior PDFs from combining the neutron-skin thickness data [$\Delta r_{np}=0.15$ fm (upper panels) or $\Delta r_{np}=0.20$ fm (lower panels) with error bars 0.02 and 0.06 fm] and the IVGDR data in $^{208}$Pb for $m_v^\star/m$ (left), $L$ (middle), and $E_{sym}^0$ (right).} \label{fig4}
\end{figure}

We start from showing in Fig.~1 the correlated posterior PDFs of isovector interaction parameters from imagined neutron-skin thickness data or those from the IVGDR data in $^{208}$Pb, after integrating the third isovector interaction parameter. It is remarkable to see that $L$ and $E_{sym}^0$ are anticorrelated with each other from the neutron-skin thickness data, while they are positively correlated from the IVGDR data. For more discussions on this behavior, we refer the reader to the appendix of Ref.~\cite{Xu20b}. It is interesting to see that the PDFs of $L$ are mostly independent of $m_v^\star$ from the neutron-skin thickness data, while those of $m_v^\star$ are mostly independent of $L$ and $E_{sym}^0$ from the IVGDR data, except for the cut off due to the limited prior distribution ranges of $L$ and $E_{sym}^0$, as can be understood from their correlation within the limited ranges in panel (i). The posterior PDFs of $m_v^\star/m$, $L$, and $E_{sym}^0$ from the imagined neutron-skin thickness data with different error bars or those from the IVGDR data in $^{208}$Pb are displayed in Fig.~2. It is seen that the neutron-skin thickness alone is unable to constrain $m_v^\star$ or $E_{sym}^0$ but only constrains $L$ within $35^{+19}_{-26}$ and $74^{+17}_{-25}$ MeV at $68\%$ confidence level around its maximum {\it a posteriori} (MAP) values, for $\Delta r_{np}=0.15 \pm 0.02$ and $0.20 \pm 0.02$ fm, respectively. However, the PDFs of $L$ become broader if the error bar changes from 0.02 fm to 0.06 fm, showing a less constraining power in the latter case. On the other hand, the IVGDR data alone is unable to constrain $L$ or $E_{sym}^0$ simultaneously but constrains $m_v^\star/m$ within $0.79\pm0.04$ at $68\%$ confidence level around its MAP value, consistent with that obtained in Ref.~\cite{Xu20c}.

Although the neutron-skin thickness data or the IVGDR data can individually pin down only one isovector interaction parameter, the data put correlations between other parameters based on the Bayesian approach. Combining both the neutron-skin thickness and the IVGDR data in $^{208}$Pb helps to constrain significantly all three isovector interaction parameters $m_v^\star$, $L$, and $E_{sym}^0$, as shown in Figs.~3 and 4. It is seen that incorporating the IVGDR data slightly reduces the MAP value of $L$ PDF from about 35 MeV to 31 MeV for $\Delta r_{np}=0.15 \pm 0.02$ fm, and from about 74 MeV to 55 MeV for $\Delta r_{np}=0.20 \pm 0.02$ fm, respectively. On the other hand, incorporating the neutron-skin thickness data has almost no effect on the PDF of $m_v^\star$. Since $L$ and $E_{sym}^0$ are correlated, $E_{sym}^0$ is also constrained around the MAP value of about 30 MeV from $\Delta r_{np}=0.15 \pm 0.02$ fm and IVGDR data, but larger than 35 MeV from $\Delta r_{np}=0.20 \pm 0.02$ fm and IVGDR data, if the range of $E_{sym}^0$ is enlarged. Again, the neutron-skin thickness data with a larger error bar of 0.06 fm do not change the PDF of $m_v^\star$ but lead to broader PDFs of both $L$ and $E_{sym}^0$. The study calls for real data of the neutron-skin thickness in $^{208}$Pb from a more reliable experimental measurement.

To summarize, we have studied the constraints on the isovector interaction parameters, i.e., the symmetry energy at the saturation density $E_{sym}^0$, the slope parameter of the symmetry energy $L$, and the isovector nucleon effective mass $m_v^\star$, from the imagined neutron-skin thickness data and the real isovector giant dipole resonance (IVGDR) data in $^{208}$Pb based on the Bayesian approach. Although the neutron-skin thickness data can only constrain $L$ and the IVGDR data can only constrain $m_v^\star$, they put correlations between other parameters. Combining both the neutron-skin thickness data and the IVGDR data helps to constrain significantly all three isovector interaction parameters.

The present study is based on the Skyrme-Hartree-Fock model and the random-phase approximation method. One sees that the correlation between model parameters can be built with given experimental data, e.g., the neutron-skin thickness and IVGDR, while the Bayesian analysis serves as a good tool to reveal that correlation. A slight different correlation could be obtained from a different theoretical model, e.g., the relativistic mean-field model, etc. It is of great interest to investigate the model dependence of such study and use additional experimental data to give a more robust constraint of model parameters by employing the Bayesian analysis.

%\begin{acknowledgments}

JX was supported by the National Natural Science Foundation of China under Grant No. 11922514. Helpful discussions with Bao-An Li and Wen-Jie Xie are acknowledged.

%\end{acknowledgments}

\end{document}